\documentclass[doublecol]{epl2} 

\usepackage{amsmath,amssymb}
\usepackage[active]{srcltx}
\usepackage{graphicx}
\usepackage{dcolumn}
\usepackage{bm}

\def\gsim{\lower.35em\hbox{$\stackrel{\textstyle>}{\textstyle\sim}$}}

\title{Graphene plasmons and retardation: strong light-matter coupling}
\shorttitle{Graphene plasmons and retardation} 

\author{G. G\'omez-Santos \and T. Stauber}

\institute{                    
  Departamento de F\'{\i}sica de la Materia 
Condensada  and Instituto Nicol\'as Cabrera,
 Universidad Aut\'onoma de Madrid, E-28049 Madrid, Spain
}

\abstract{We study the retardation regime of doped graphene plasmons,
  given by the nominal crossing of the unretarded plasmon and
  light-cone. In addition to modifications in the plasmon dispersion
  relation, retardation implies strong coupling between propagating
  light and matter, even for homogeneous graphene, which opens up the
  possibility of efficient plasmonics in simple graphene devices. We
  exemplify this enhancement in a double-layer configuration that
  exhibits {\em perfect} (if lossless) light transmissions across a
  classically forbidden region, providing a simpler analog of the
  corresponding phenomenon in perforated metal sheets. We also show
  that (broad) Fabry-P\'erot resonances present without graphene turn
  into sharply peaked, quasi-discrete modes in the presence of
  graphene where graphene's response function is given by the typical
  Fano lineshape.}

\begin{document}

\maketitle

\section{Introduction}\label{intro}
In addition to its amazing transport and mechanical 
properties\cite{Geim09}, graphene's
optical behavior is also notable. 
Absorption, for instance, has the experimentally observed\cite{Nair08,Mak08},
universal value $\approx \pi \alpha$ for light in the visible spectrum,
depending on the fine structure constant $\alpha$, but not on material's
properties. 
Several proposals and/or realizations highlight graphene
 potential  in optical and communication
technologies, including resistive touchscreens of transparent and flexible
displays,\cite{Bae10,Bonaccorso10}, opto-electronic devices such as 
photodetector\cite{Mueller10}, broadband
absorber\cite{Liu11},  mode-locked laser due to current
saturation\cite{Sun10}, 
or as a highly efficient fluorescence 
quencher\cite{Treossi09,Kim10,Chen10,Loh10,Swathi09,Velizhanin11,G-S11}.

Doped graphene has attracted much attention recently as a suitable candidate
for noble metal's replacement in the active field of 
plasmonics\cite{Jablan09,Vakil11,Koppens11,Nikitin11}. In addition to
their intrinsic importance as a probe for dynamics,  
graphene plasmons\cite{Wunsch06,Hwang07,Barlas07} (GPs)
offer a number of advantages such as frequency tunability, long life-times, and
spatial confinement due shorter (than light) wavelengths.  They have been
observed via electron energy loss spectroscopy\cite{Seyller08,Tegenkamp11} and
near-field nanoscopy.\cite{Basov11,Chen12,Fei12}

 Unfortunately, GPs  do not couple easily to propagating electromagnetic modes
over most of its range due to the large momentum mismatch,  complicating their
manipulation. This problem could be overcome by breaking the conservation of
parallel momentum either with confined geometries\cite{Nikitin11,Christensen12}
or artificial periodicities\cite{Ju11,Echtermeyer11,Thongrattanasiri12}. However, there is a regime where even homogeneous
GPs must couple strongly to (propagating) light: the retardation limit. It
applies for frequencies below a characteristic crossover scale, which can be
taken as the crossing between the nominal, unretarded GP dispersion and the
light-cone. There,  phenomena associated with strong light-graphene coupling
take place.    

Here, we demonstrate this enhanced light-graphene coupling and apply it to
a double-layer graphene\cite{Hwang09,Stauber12} arrangement possessing {\it extraordinary transmission}. This
term was originally coined to describe the enormous transmission experimentally
observed through periodically perforated metal
sheets\cite{Ebbesen98,Moreno01,Vidal10}, a situation where the naive
expectation would have prescribed just the opposite. An explanation was
provided in terms of the excitation of  surface plasmons, with the result of
enhanced transmission ({\em perfect} in the absence of absorption) through a
nominally opaque region. In our case, the resonant coherent excitations
of the graphene layers allow the enhanced
transmission of photons through the central, classically forbidden region for
photons, in direct analogy with the metallic case.
In the propagating region,  the strongly quenched transmission turns into
perfect transmission at sharply peaked resonances related to the original
Fabry-P\'erot resonances, where graphene's response function is given by the
typical Fano lineshape due to the nearly discrete nature of the spectrum in the
central slab.

The Letter is organized as follows. We first study the retardation limit  and its effect on the GP dispersion for  single-layer graphene. Then we consider the retardation regime for double-layer graphene separating different dielectric media where extraordinary transmission takes place. Finally we summarize our results and an appendix details the GPs dispersion relation for the double-layer arrangement in the retardation limit.

\section{Retardation limit. Single-layer graphene}\label{single}

The standard expression for a 2d plasmon, $ \omega_p \propto \sqrt{q}
$, assumes instantaneous Coulomb coupling between
charges.\cite{Wunsch06,Hwang07} Therefore, it cannot be correct when
the nominal plasmon dispersion meets the light-cone, $\omega_p
\lesssim cq $. This is the regime where retardation effects matter and
strong light-graphene interaction takes place, as we now show.

 Consider a graphene layer sandwiched between two dielectrics with  permittivities
$\varepsilon_{1,2} = \epsilon_{\scriptscriptstyle{1,2}}\, \varepsilon_0$ 
and $\epsilon_1>\epsilon_2$. 
 The longitudinal current response to an (in-plane) longitudinal, 
 external vector potential is given by the standard RPA expression
\begin{equation}\label{RPA}
\chi_l = \frac{\chi_l^o} {1 - e^2 d_l\chi_l^o }
,\end{equation}  
where $ \chi_l = {\cal G}_{jj}$ is the retarded Green function for the
longitudinal current,  $\chi_l^o$ its non-interacting (bare) version,   and
$d_l={\cal G}_{AA} $ that of the (in-plane) longitudinal vector potential (in
the absence of graphene).
For graphene, we use the well-known results\cite{Wunsch06,Hwang07,Stauber10}
 for $\chi_l^o$, whereas for the photon field, one
straightforwardly finds\cite{G-S11} $ d_l = \tfrac{q'_1 q'_2 \,\omega^{-2}}{\varepsilon_2 q'_1+ \varepsilon_1 q'_2}$, with 
$ q'_i = \sqrt{q^2 - (\omega/ c)^{2} \epsilon_i }$.

GPs are the poles of Eq. \ref{RPA}, leading to the following  
(implicit)  dispersion relation,
including retardation

\begin{equation}
\omega^2 = e^2 \frac{q'_1q'_2}{\varepsilon_2 q'_1+ \varepsilon_1 q'_2} \chi_l^o 
.\end{equation}   

\begin{figure} 
\includegraphics[clip,width=8cm]{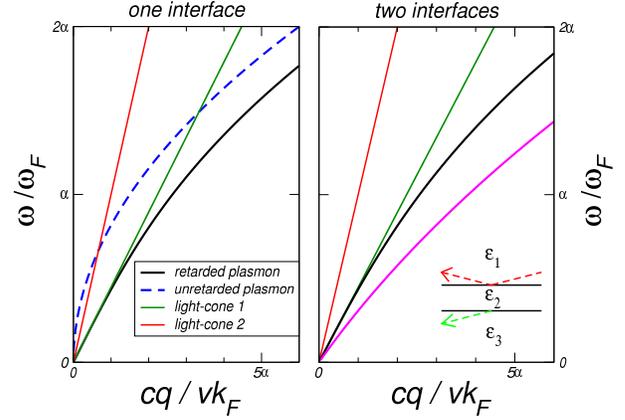} \\  %
\caption{Left panel:
Longitudinal plasmon exact dispersion relation (black continuous line), compared
to the instantaneous approximation (blue dashed line) for doped graphene between
medium 1 ($\epsilon_1=5$) and medium 2 ($\epsilon_2=1$), light-cones also shown.
Right panel: As in left panel for the double-layer graphene showing the 
in-phase (upper curve, black) and out-of-phase (lower curve, magenta) 
plasmons for $\epsilon_3=\epsilon_1$ and layer separation 
$\tilde{z} = \tilde{z}_c/2$. 
Inset:  schematic geometry and  transmission setup.} 
\label{retplsmnfig}
\end{figure}

In the unretarded limit, $c \rightarrow \infty $, this gives the  known
square-root dispersion. In contrast, the exact solution replaces this behavior
with a linear dispersion which merges with the slower medium (1) light-cone
below a characteristic crossover frequency, as illustrated in 
Fig. \ref{retplsmnfig} (left panel), and summarized as follows:
\begin{equation}\label{dispersion} 
\left( \dfrac{\omega}{\omega_F}\right)^2 = \left \{ 
\begin{array}{ll}
\left(\frac{4 \alpha_g}{\epsilon_1 + \epsilon_2}\right) \left(\frac{q}{k_F}\right),  & \omega
\gtrsim \omega_c \\ \;&\\ \left(\frac{c_1}{v}\right)^2 \left(\frac{q}{k_F}\right)^2, & \omega
\lesssim \omega_c  
\end{array}
\right. 
,\end{equation}    
where $\omega_F,k_F$ and $v$ are graphene's Fermi  frequency, Fermi momentum
and Fermi velocity, $c_1=c/\sqrt{\epsilon_1}$ is  medium 1 (slower) light velocity, and 
$\alpha_g = \tfrac{c}{v} \alpha$ represents graphene fine structure constant.
The crossover between regimes  takes place for frequencies  which roughly
corresponds to the intersection of the unretarded GP and light-cone
dispersions.  This scale is given by    $\omega_c \sim \alpha \omega_F $ for
reasonable  dielectric constants. Typical frequencies are $\nu_c \sim 200
\,\text{GHz} $  for doping level  $n \sim 10^{12} \,\text{cm}^{-2}$, $\nu_c \sim 600
\,\text{GHz} $  for doping level  $n \sim 10^{13} \,\text{cm}^{-2}$, reaching
the technologically important  $\text{THz}$ regime  for  $n \sim 10^{14}
\,\text{cm}^{-2}$. 

Therefore, for $ \omega \lesssim \omega_c $ retardation always matters. Not surprisingly, 
this linear regime is also the region of strong graphene-light coupling. This is immediately
seen by looking at the reflection and transmission amplitudes for the (in-plane) longitudinal
vector potential upon passing from medium $i$ to $j$, given by 
\begin{equation}\label{eqrij} 
\begin{split} 
r_{ij} & = 
\frac{\varepsilon_i q'_j - \varepsilon_j q'_i + e^2 q'_i q'_j  \chi_l^o \omega^{-2}} {\varepsilon_i q'_j
+ \varepsilon_j q'_i - e^2 q'_i q'_j  \chi_l^o \omega^{-2}}\\ 
t_{ij} & = 1 + r_{ij} 
\end{split}
.\end{equation}   
For interband transitions at frequencies above $2 \,\omega_F $, graphene
terms in Eq. \ref{eqrij} are  minute, leading to the universal $2.3 \%$ weak
absorption in vacuum, for instance. On the other hand, for frequencies below
$\omega_c $, graphene response  starts to dominate in Eq. \ref{eqrij} implying
strong radiation-graphene coupling.  
For instance, in the limit $\omega \ll \omega_c $ 
the   reflection amplitude becomes 
\begin{equation}\label{small} 
r = -1 + {\cal O}(\frac{\omega}{\omega_c})
.\end{equation}   
 $r = -1 $ implies perfect reflection or, equivalently, a zero photon field at
graphene position as a boundary condition. Therefore,  the (small) parameter 
$ \frac{\omega}{\omega_c} \sim \frac{\omega}{\alpha \omega_F}$ of our strong
coupling regime also measures the
departure from this zero-field boundary condition.

 \section{Double-layer graphene. Extraordinary transmission}\label{double}

 We consider two identically doped graphene sheets, separating three 
 dielectrics with permittivities  $ \epsilon_{1,2,3}$, see Fig.
 \ref{retplsmnfig} (right panel), and impose left-right symmetry choosing
 $\epsilon_2 < \epsilon_1  = \epsilon_3$. Under these conditions, for large
 enough incident angles the central region does not support propagating modes:
 light must {\it tunnel} through this evanescent region. Doped graphene
 enhanced response in the retardation regime  can make this tunneling {\em
 perfect}, as we will show.  

 Reflection and transmission coefficients can be obtained from the corresponding
 amplitudes,  written as 

\begin{equation}\label{tilderij}
\begin{split}
\tilde r_{1,3} & = r_{12} + \frac{t_{12} r_{23} t_{21} 
\text{e}^{-2 q'_2 z}}{1 -r_{21} r_{23} \text{e}^{-2 q'_2 z} } \\
\tilde t_{1,3} &= \frac{t_{12} t_{23}\text{e}^{ -q'_2 z}}{1 -r_{21} r_{23} \text{e}^{-2 q'_2 z}}
\end{split}
,\end{equation}  
for light incoming from medium $1$ and being transmitted to medium $3$ 
(see  Fig. \ref{retplsmnfig} inset), with $t_{ij}$ and $r_{ij}$ taken from Eq.
\ref{eqrij}. 

In order to display the connection between light propagation and graphene
behavior, we also calculate the double-layer (longitudinal) current-current
response, given by the following matrix generalization of Eq. \ref{RPA}  %
\begin{equation}\label{RPAmatrix}
\bm \chi_l = (\bm 1 - e^2 \bm \chi_l^o \bm d_l)^{-1} \bm \chi_l^o 
,\end{equation}  
with the $2\text{x}2$ matrix $\bm \chi_l^o = \text{diag}(\chi_1^o,\chi_2^o)$
representing the non-interacting (longitudinal) response of graphene layers $1$
and $2$, whereas the photonic matrix  $\bm d = (d_{ij})$ is given by
$d_{11}=d_1\,(1+\tilde{r}^o_{1,3})$,  $d_{22}=d_3\,(1+\tilde{r}^o_{3,1})$, and 
$d_{12}=d_1\,\tilde{t}^o_{1,3}=d_3\,\tilde{t}^o_{3,1}=d_{21}$.  Here,
$d_1=\tfrac{q'_1}{2 \varepsilon_1 \omega^2}$,  $d_3=\tfrac{q'_3}{2 \varepsilon_3
\omega^2}$, and  $\tilde{r}^o_{i,j} (\tilde{t}^o_{i,j})$ correspond to the
expressions of Eq. \ref{tilderij} evaluated for the dielectric geometry of Fig.
\ref{retplsmnfig}, but without graphene layers ($\chi_l^o = 0$ in 
Eq. \ref{eqrij}).
For our left-right symmetric arrangement, the in-phase and out-of-phase
components of graphene currents, $j_{\pm}=j_1\pm j_2$, diagonalize $\bm
\chi_l$, with corresponding (complex) eigenvalues that we denote by 
$\chi_{\scriptscriptstyle{++}}$ and $\chi_{\scriptscriptstyle{--}}$,
respectively.

Genuine (non radiative) GPs correspond to the poles of $\bm \chi_l$ in the totally evanescent
regime, that is, the region to the right of the slower light-cone in Fig.
\ref{retplsmnfig}. In the Appendix we show that there always are two GPs, one
for each component of the response.  The in-phase GP 
$(\chi_{\scriptscriptstyle{++}})$ merges with the corresponding limiting
light-cone, whereas the out-of-phase  GP $(\chi_{\scriptscriptstyle{--}})$
either merges with the light-cone or develops its own linear dispersion at
slower velocity if the separation between layers is below a critical value
$\tilde{z}_c=z_c\,k_F$, with 
$\tilde{z}_c=\tfrac{\epsilon_2}{\epsilon_1-\epsilon_2} \tfrac{c}{2\,v\,\alpha}
$. Both are shown in  Fig. \ref{retplsmnfig} (right panel) where we have used
$\tilde{z}=\tilde{z}_c/2$, and permittivities $\epsilon_2=1$ (vacuum) and
$\epsilon_1=\epsilon_3 =5$, a representative  value for common substrates. For
bulk BN, we have, e.g., $\epsilon_{1,3}\sim5$, whereas  $\epsilon_{1,3}\sim 4$
for few layer BN or SiO${}_{2}$.

\begin{figure} 
\includegraphics[clip,width=8cm]{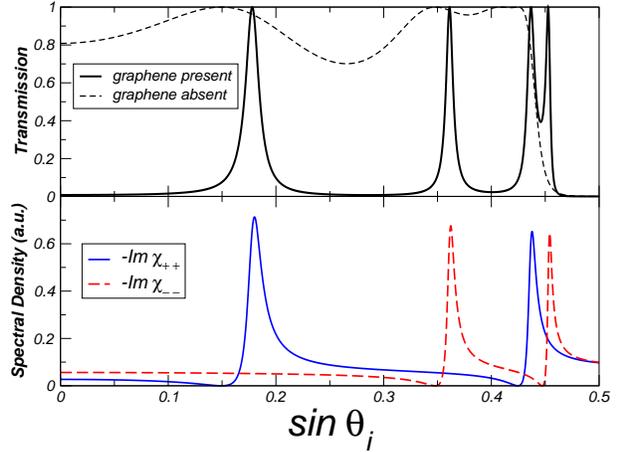} \\ 
\caption{Top panel:
 Transmission as a function of incoming angle for frequency 
$\omega = (\alpha/2) \, \omega_F$ and  double-layer graphene separation 
$ z k_F = 20 \, c / (v\,\alpha)$ for permittivities  $ \epsilon_{1,3} = 5,\;\epsilon_2 = 1$.
Dashed (black) line: same as before in the absence of graphene.
Bottom panel: 
 Spectral density of the in-phase (solid, blue line)
and out-of-phase (dashed,  red line) double-layer 
 graphene response.} 
\label{fabry}
\end{figure}

Though GPs remain in the evanescent region of both media, the presence of
graphene  also strongly affects propagating light, leading to
resonances with perfect transmission. At first glance, this statement seems
incompatible with the (almost) perfect reflective character of a graphene layer 
in our strong coupling regime. Yet, the zero field boundary condition also
decouples the photon field in the central slab from the propagating modes
outside, leading to the standard discrete spectrum for a particle in a box.
Now, the small parameter 
$ \frac{\omega}{\omega_c} \sim \frac{\omega}{\alpha \omega_F}$ measuring the departure from perfect reflection is
also a measure of the (small) coupling between the discrete modes of the
central slab and the propagating modes outside, leading to (radiative) resonances.
Drastic changes are expected close to these resonances and, indeed, we will see
that the otherwise expected strong reflection turns into perfect transmission.

These
resonances are shown in Fig. \ref{fabry} (top panel), where we present the
transmission for layer's separation $z\,k_F = 20\,c/(v\alpha)$ and
frequency $\omega = (\alpha/2)\,\omega_F$. For a doping level of $n
\sim 10^{12}\text{cm}^{-2}$, this would imply a frequency $\nu \sim
100 \;\text{GHz}$, and separation $ z \sim 5\times10^{-3}
\text{m}$. For  $n \sim 10^{14}\text{cm}^{-2}$, we have  $\nu \sim  1\;\text{THz}$, and 
 $z \sim 5\times10^{-4} \text{m}$.
Although  the existence of perfect transmission at Fabry-P\'erot-like
resonances is 
generic for this geometry even without graphene 
(see dotted line in top panel of Fig.\ref{fabry}),
these resonances are  sharply modified by its presence. The otherwise strongly
reflective graphene confines the transmission to narrow resonances close to the
region where the almost discrete central slab modes are excited. The narrow
width of these resonances is a measure of the long lifetime of the central
photonic modes, weakly decaying into the radiative modes outside. A direct
evidence of the role of these (almost) discrete spectrum is provided by the   
spectral densities of the in-phase
$(\chi_{\scriptscriptstyle{++}})$ and out-of-phase
$(\chi_{\scriptscriptstyle{--}})$ graphene  response, also shown
in Fig. \ref{fabry} (bottom panel). Up to an overall scale, these spectral
densities coincide with the spectral densities of the photonic field and,
therefore, the sharp resonances in graphene match the expected sharp spectral
densities of weakly decaying, almost discrete photonic modes.
Notice the  one-to-one
correspondence between response and  transmission peaks, emphasizing
 the role of these modes in overturning the otherwise generic strong reflection
 of graphene in the strong coupling limit.


 The spectral functions of the bottom panel of Fig. \ref{fabry} exhibit a marked asymmetric
Fano-like profile, vanishing precisely where grapheneless transmission is one.
This Fano-like profile can be expected, given the key role played by the almost
discrete modes of the central slab. It can be quantitatively explained noticing
that 
 \begin{equation}\label{fano1} \chi_{\scriptscriptstyle{\pm\pm}} \sim 
\frac{1}{ 1 - e^2 \chi_l^o  (d_{11} \pm d_{12})}  ,\end{equation}   
  but the
photon propagator in the absence of graphene vanishes at the  corresponding 
grapheneless  resonances as  
\begin{equation}\label{fano2} 
d_{11} \pm d_{12}
\sim a (\delta s) + \text{i} b(\delta s)^2 ,\end{equation}
where $a$ and $b$ are real constants, and $\delta s $ is the departure of
$sin(\theta_i)$ from the  value corresponding to the  discrete modes upon
strictly enforcing zero-field boundary conditions, given by   $\exp(-2 q'_2 z) =
1$.   Notice that this zero-field boundary condition for discrete modes in the
presence of graphene  coincides with the perfect transmission  condition in the
absence of graphene, explaining the vicinity of both points.  The small shift of
the actual solution being the consequence of  the  small leaking of  discrete
modes into the outside continuum. This  leaking  also accounts  for the small
width of the resonance, and leads to a characteristic resonance asymmetry, given
by  
$
-\text{Im} \chi_{\scriptscriptstyle{\pm\pm}} \sim \frac{ (\delta
s)^2}{(\delta s-s_*)^2 + l^2 (\delta s)^4},$ 
%
where $s_*  \sim \frac{\omega/\omega_F}{\alpha l} $ is the resonance  shift from the nominal discrete
modes graphene, playing the role of small parameter in the strong coupling
regime, and  $l$  
being (up to a factor  of order one) the central slab width 
in units of the minimum required for the emergence
of the first nominal discrete mode. This expression can be approximately recast 
 in the more common Fano lineshape\cite{Fano61} ,
\begin{equation}\label{fano4}
-\text{Im} \chi_{\scriptscriptstyle{\pm\pm}} \sim \frac{(Q\gamma/2 + \delta
s')^2}{(\delta s')^2 + (\gamma/2)^2} 
,\end{equation} 
with the identifications $\delta s' = \delta s-s_* $, the width 
$\gamma\sim l s_{*}^2$, and the Fano asymmetry parameter 
$Q=\tfrac{2 s_*}{\gamma}$
with value $Q \sim 2-3$, for fig. \ref{fabry} (bottom panel).

\begin{figure} 
\includegraphics[clip,width=8cm]{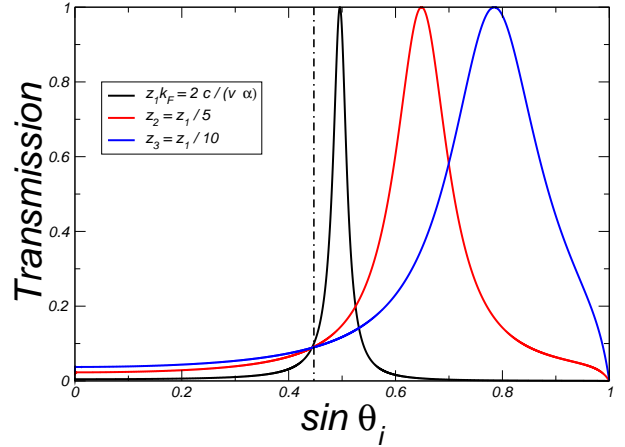} \\ 
\caption{Transmission as a function of incoming angle for frequency 
$\omega = (\alpha/2) \, \omega_F$, permittivities  $ \epsilon_{1,3} = 5,\;\epsilon_2 = 1$, 
and double-layer graphene  separations:  
$ z_1 k_F = 2 \, c / (v\,\alpha)$ (black),  
$ z_2  = z_1 / 5$ (red), and 
$ z_3  = z_1 / 10$ (blue). 
The vertical dashed line marks the onset of the evanescent regime in
 the central slab.
 } 
\label{perfect}
\end{figure}

\begin{figure} 
\includegraphics[clip,width=8cm]{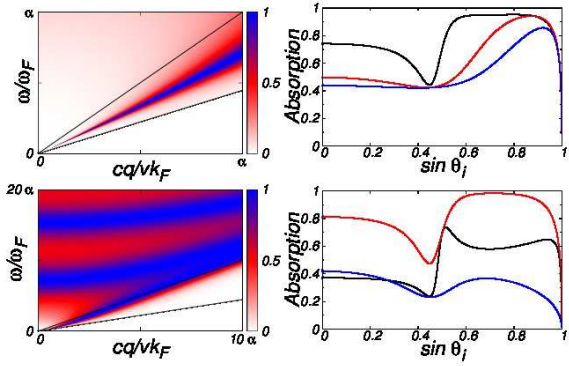} \\ 
\caption{
Top left panel: Transmission in the $\omega-q$ plane showing the perfect
resonance within the evanescent region for double-layer separation corresponding
to $z_2 k_F= 2 \, c / (5 v\,\alpha)$.
Bottom left panel: As in top left panel for an extended region of the 
$\omega-q$ plane to include additional Fabry-P\'erot-like resonances.
Top right panel:  Absorption when graphene losses are included with 
$\omega \tau = 0.65$ for the same double-layer separations (and colors)  
as in Fig. \ref{perfect}. 
Bottom right panel:  Absorption for fixed double-layer separation
$z_1 k_F= 2 \, c / (v\,\alpha)$ and graphene losses 
$\omega \tau = 2$ (black), 
$\omega \tau = 0.5$ (red), and 
$\omega \tau = 0.05$ (blue).} 
\label{fig4}
\end{figure}

Interestingly, the rightmost resonance in Fig. \ref{fabry} occurs for  $  \sin
\theta_i > 1/\sqrt{5} \sim 0.447$, that is, in the evanescent regime for  medium
$2$. 
In fact, we can tune the arrangement to make this last resonance the
only one present. This is shown in Fig. \ref{perfect}, where we present the
transmission  calculated with  the same frequency as before  $\omega =
(\alpha/2) \omega_F$, but for three shorter separations given by 
$z_1\,k_F = 2\,c/(v\alpha)$, $z_2 = z_1/5$, and $z_3 = z_1/10$ 
     $( \nu \sim 100 \,\text{GHz}\; \text{and}\;z_1 \sim 5\times10^{-4} \text{m}\; \text{for}\;
 n \sim 10^{12}\,\text{cm}^{-2})$. 
   One can observe a  perfect transmission peak within the 
classically forbidden region for photons. This resonance is rather sharp close
to the propagating boundary, moving deeper into the evanescent region and broadening
with decreasing double-layer separation. Notice that, for the cases shown in Fig.
\ref{perfect}, the transparency window is virtually confined to the evanescent
regime.   
Although a finite photon tunneling
always exists, the fact that the transmission becomes perfect in the evanescent
region\cite{GdeAbajo05} is entirely due to the presence of doped graphene
sheets, and  
is 
 associated with a resonance in their out-of-phase response (see bottom panel of
 Fig. \ref{fabry}). This 
perfect transmission peak within the evanescent region extends all the way down
to zero frequency, as shown in the top left panel of Fig. \ref{fig4}. 
For higher
frequencies, this resonance approaches the upper light-cone and  
merges with the lowest propagating Fabry-P\'erot like
resonance, as seen in the bottom  left panel of Fig. \ref{fig4}. 

Our aim so far has been to expose the strong light-graphene coupling
of the retardation regime in a simple manner. Therefore, we have
ignored absorption.  Graphene losses will degrade the perfect nature
of transmission, making the possible observation of coherent effects
more likely in reflection rather than transmission. Although we are
not aware of any intrinsic limitation in the minimization of losses
attainable in doped graphene, the low frequencies of the
strong-coupling regime might pose a severe handicap.  Nevertheless,
absorption opens new possibilities: the sharp response of graphene in
the absence of absorption suggests a correspondingly enhanced
absorption when losses are allowed. This is the case as the upper right
panel of Fig. \ref{fig4} shows, where the absorption for a lifetime
$\tau\sim 10^{-12}s$, present in suspended graphene and corresponding to
$\omega\tau=0.65 $, is plotted for the same double-layer separations
of Fig.  \ref{perfect}. Indeed, one observes an overall high
absorption, reaching values above $90\%$ precisely in the evanescent
region.  From this perspective, our arrangement is closely connected
to that of Ref.\cite{Bludov10}, providing a complementary and
potentially simpler alternative to absorption enhancement based on
periodic patterning\cite{Ju11,Echtermeyer11,Thongrattanasiri12}. It is
similar, though, to the previously suggested enhanced absorption of
graphene placed in a (double) Fabry-P\'erot
cavity.\cite{Furchi12,Ferreira12}

In the bottom right panel of Fig. \ref{fig4}, the absorption for
different intrinsic damping rates corresponding to
$\omega\tau=2,0.5,0.05$ is shown where the last value would assume a
phenomenological relaxation rate $\gamma$ of the polarization with
$\hbar \gamma\sim10$meV, present in graphene on a
SiO$_2$-substrate. Notice that the double-layer absorption can display
a non-monotonous behavior with the intrinsic single-layer graphene
losses, providing a richer degree of control.

\section{Summary}\label{summ}

We have studied the retardation regime of doped graphene plasmons,
given by $ (\omega/\omega_F) \lesssim \alpha$. Apart from modifying
the unretarded GP dispersion behavior from $\sqrt{q}$ to $q$, such a
limit marks the onset of strong radiation-graphene coupling and we
have exhibited such enhancement in the simplest double-layer
arrangement. i) We first observed that the transmission is strongly
quenched except at sharply peaked resonances,
consistent with the nearly discrete spectrum of the strongly confined
 field in the central slab brought about by graphene, whose
 response function is given
by the typical Fano lineshape. ii) We
also discussed that the transmission, perfect if losses are ignored,
can be confined to the classically forbidden region for photons
between graphene layers, providing a direct analog of the
extraordinary transmission through perforated metals in a conceptually
simpler setup. iii) Finally, we showed that enhanced transmission becomes
increased absorption when losses are allowed, displaying
non-monotonous behavior. Graphene's doping tunability in this or similar
setups should thus open up new ways to efficiently control the flow
and/or absorption of radiation.

\acknowledgments
This work has been supported by FCT via grant PTDC/FIS/101434/2008 and by MICINN via grant FIS2010-21883-C02-02.

\section{Appendix} We outline here the calculation of double-layer GPs  in the retarded regime.
We take $\epsilon_3 = \epsilon_1 > \epsilon_2$. Therefore, genuine GPs are the response poles
located to the right of the slower (outer media) light-cone. These are given by the solutions of
\begin{equation}\label{pm}
1  \mp r_{21} \text{e}^{- q'_2 z}  = 0 
,\end{equation}  
corresponding to the in-phase $(\chi_{{\scriptscriptstyle++}})$ and out-of-phase $(\chi_{\scriptscriptstyle{--}})$ response
components, respectively. Taking $r_{21}$  from Eq. \ref{eqrij}  deep into the retarded
regime, one finds the in-phase GP as
\begin{equation}
\epsilon_1 \tilde{\omega}^2_{\scriptscriptstyle+} \rightarrow  
4 \pi \alpha_g \tilde{\chi}^o \tilde{q}'_1
,\end{equation}  
where we have used dimensionless magnitudes: $\tilde{\omega}={\omega}/{\omega_F}$, 
$\alpha_g = \tilde{c} \alpha$, 
$ \tilde{c} = c/v$,  $ \tilde{c}_i^2 = \tilde{c}^2/\epsilon_i $, $\tilde{q}=q/k_F$,  $ \tilde{q}'_i = \sqrt{\tilde{q}^2-\tilde{\omega}^2 /\tilde{c}_i^2}$,
 with  single-layer graphene response,
$\tilde{\chi}^o=\pi^{-1} $,
 taken in the local approximation, excellent for the retarded regime. In
this limit,
 $\tilde{\omega}_{\scriptscriptstyle+}$ 
 coincides with the GP pole of an isolated graphene
layer in the boundary between two (semi-infinite) media $1$ and $2$. Therefore, one easily
sees that it merges
asymptotically with  the outer media light-cone, as in Eq. \ref{dispersion}:
\begin{equation}
\tilde{\omega}_{\scriptscriptstyle+} = \tilde{c}_1 \tilde{q} + \mathcal{O} (\tilde{q}^3)
.\end{equation}  

 Analyzing the {\em plus} Eq. \ref{pm} for the out-of-phase GP, 
a critical layer separation emerges, $ z_c$, given by
\begin{equation}\label{zc}
z_c \,k_F = \tilde{z}_c = \frac{\epsilon_2}{2 (\epsilon_1-\epsilon_2)}
\frac{\tilde{c}}{\alpha}
,\end{equation}  
that separates the following two regimes  
\begin{equation} 
 \tilde{\omega}_{\scriptscriptstyle-} = \left \{ 
\begin{array}{ll}
\tilde{c}_1     \tilde{q} + \mathcal{O} (\tilde{q}^3),     & \tilde{z} > \tilde{z}_c \\
\tilde{v}_{\scriptscriptstyle-} \tilde{q} + \mathcal{O} (\tilde{q}^3),     & \tilde{z} < \tilde{z}_c
\end{array}
\right. 
,\end{equation}    
where
\begin{equation}
\tilde{v}_{\scriptscriptstyle-}^2 = \frac{\tilde{c}_1^2}{1 + (1 - \frac{\epsilon_2}{\epsilon_1}) 
(\frac{\tilde{z}_c}{\tilde{z}}-1)}
.\end{equation}
We conclude that the out-of-phase GP develops a linear dispersion relation that either merges
with the outer media light-cone (and, therefore, with $\tilde{\omega}_{\scriptscriptstyle+}$ from below) for
$\tilde{z}>\tilde{z}_c$,  or shows a lower velocity $\tilde{v}_{\scriptscriptstyle-} < \tilde{c}_{1}$ for
$\tilde{z}<\tilde{z}_c$, as stated in the paper. 

\bibliographystyle{eplbib}
\providecommand{\noopsort}[1]{}\providecommand{\singleletter}[1]{#1}

\end{document}